\documentclass[prd,aps,showpacs,tightenlines,nofootinbib,preprint,preprintnumbers]{revtex4}
\usepackage{graphicx}
\usepackage{amsfonts}
\usepackage{amssymb}
\usepackage{latexsym}
\newcommand{\CO}{{\cal O}} 
 \newcommand{\CR}{{\cal R}}

\newcommand{\bear}{\begin{array}}  \newcommand{\eear}{\end{array}}
\newcommand{\bea}{\begin{eqnarray}}  \newcommand{\eea}{\end{eqnarray}}
\newcommand{\beq}{\begin{equation}}  \newcommand{\eeq}{\end{equation}}
\newcommand{\bef}{\begin{figure}}  \newcommand{\eef}{\end{figure}}
\newcommand{\bec}{\begin{center}}  \newcommand{\eec}{\end{center}}
\newcommand{\non}{\nonumber}  
\newcommand{\lmk}{\left(}  \newcommand{\rmk}{\right)}
\newcommand{\lkk}{\left[}  \newcommand{\rkk}{\right]}
\newcommand{\lhk}{\left \{ }  \newcommand{\rhk}{\right \} }
\newcommand{\lnk}{\left \{ }  \newcommand{\rnk}{\right \} }

\newcommand{\bib}{\bibitem}

\newcommand{\ns}{n_s}
\newcommand{\Mpc}{{\rm Mpc}}
\newcommand{\mg}{M_G}

\newcommand{\Psibar}{\overline{\Psi}}
\newcommand{\muni}{\mu^2}

\newcommand{\kn}{c_N}

\newcommand{\Phimin}{\Phi_{\min}}
\newcommand{\calr}{{\cal R}}
\def\IBIDD#1#2#3{{\it ibid}. {\bf #1}, #2 (20#3)}

\def\APJJ#1#2#3{Astrophys. J. {\bf #1}, #2 (20#3)}

\def\JL#1#2#3{JETP Lett. {\bf #1}, #2 (19#3)}

\def\JHEPP#1#2#3{J. High Energy Phys. {\bf #1}, #2 (20#3)}

\def\PLB#1#2#3{Phys. Lett. B {\bf #1}, #2 (19#3)}
\def\PLBB#1#2#3{Phys. Lett. B {\bf #1}, #2 (20#3)}
\def\PLBold#1#2#3{Phys. Lett. {\bf#1B}, #2 (19#3)}

\def\PRD#1#2#3{Phys. Rev. D {\bf #1}, #2 (19#3)}
\def\PRDD#1#2#3{Phys. Rev. D {\bf #1}, #2 (20#3)}

\begin{document}

\title{Smooth hybrid inflation in supergravity \\ with a running
spectral index and early star formation}

\author{Masahide Yamaguchi\footnote{Present address: Department of
Physics and Mathematics, Aoyama Gakuin University, Kanagawa 229-8558, Japan.}
} 
\affiliation{Physics Department, Brown University, Providence, Rhode Island
  02912, USA}
\author{Jun'ichi Yokoyama} 
\affiliation{Department of Earth and Space Science, Graduate School of
  Science, Osaka University, Toyonaka 560-0043, Japan}
%%%

\date{\today}
\preprint{BROWN-HET-1397}
\preprint{OU-TAP-226}

%\maketitle

\begin{abstract}
It is shown that in a smooth hybrid inflation model in supergravity
adiabatic fluctuations with a running spectral index with $\ns >1$ on a
large scale and $\ns <1$ on a smaller scale can be naturally generated,
as favored by the first-year data of WMAP. It is due to the balance
between the nonrenormalizable term in the superpotential and the
supergravity effect. However, since smooth hybrid inflation does not
last long enough to reproduce the central value of observation, we
invoke new inflation after the first inflation.  Its initial condition
is set dynamically during smooth hybrid inflation and the spectrum of
fluctuations generated in this regime can have an appropriate shape to
realize early star formation as found by WMAP.  Hence two new features
of WMAP observations are theoretically explained in a unified manner.
\end{abstract}

%\pacs{98.80.Cq,04.65.+e,11.27.+d \hspace{4.0cm} BROWN-HET-, OU-TAP-}

\maketitle

%\newpage
%\tighten

%%%%%%%%%%%%%%%%%%%%%%%%%%%%%%%%%%%%%%%%%%%%%%%%%%%%%%%%%%%%%%%%%%%%%
\section{Introduction}
%%%%%%%%%%%%%%%%%%%%%%%%%%%%%%%%%%%%%%%%%%%%%%%%%%%%%%%%%%%%%%%%%%%%%

\label{sec:intro}

The observation of cosmic microwave background anisotropy by the
Wilkinson Microwave Anisotropy Probe (WMAP) has determined cosmological
parameters with a good accuracy. They have shown that the geometry of
our universe is practically flat and that the energy density of the
universe is dominated by dark energy and compensated by dark matter and
a small amount of baryons \cite{Bennett,Spergel}. Furthermore,
primordial density fluctuations are shown to be adiabatic, Gaussian, and
nearly scale invariant, which suggests that they were produced during
inflation \cite{Spergel,Peiris}. Thus, the so-called concordance model
was confirmed.

Going into the details, however, we find several interesting features
that may not be reconciled with a simple scale-invariant spectrum :
namely, an early period of re-ionization (see, e.g., \cite{YY} for a
possible explanation), a lack of fluctuation power on the largest
scale,\footnote{This feature was already seen in COBE observations and a
possible explanation was proposed before WMAP data were released
\cite{JY99}.} the running spectral index of fluctuations, $\ns$
\cite{YY,FLZZ,KS,KYY}, and the oscillatory behavior of multipoles which
may suggest oscillations in the primordial power spectrum
\cite{kogo}. Of course, these properties may disappear eventually when
the observations are improved. But it is still important to consider a
model to explain them at the present stage.  In this paper, we discuss
the running of spectral index from $\ns>1$ on a large scale to $\ns<1$
on a smaller scale. More concretely, it is shown that $\ns=1.13\pm 0.08$
and $d\ns/d\ln k=-0.055^{+0.028}_{-0.029}$ on the scale
$k_0=0.002\,\Mpc^{-1}$ \cite{Peiris}.

Among the three types of slow-roll inflation--namely, new \cite{newinf},
chaotic \cite{chaoinf}, and hybrid \cite{hybrid} inflation--the first
two scenarios predict fluctuations with $\ns < 1$ while hybrid inflation
can realize both those with $\ns >1$ and $\ns < 1$.  Although it is
fairly easy to construct a model whose spectral index runs from $\ns <1$
to $\ns > 1$ for decreasing length scales, it is quite nontrivial to
realize the opposite running.  The hybrid inflation model in
supergravity proposed by Linde and Riotto \cite{LR} is an exception in
which the desired feature is realized due to the contributions to the
potential from both one-loop effects and supergravity effects.  Based on
this observation, some models have been discussed in an attempt to
reproduce the results of the WMAP, but it turned out that the large
enough running is incompatible with long enough inflation
\cite{KS,KYY,YY}.  This is because the Yukawa coupling constant must be
relatively small for sufficient inflation while it must be large for
large running.

This problem was first solved in \cite{KYY} by introducing another
inflaton whose appropriate initial condition is automatically prepared
during the hybrid inflation regime
\cite{IY,Kawasaki:1998vx,Kanazawa:1999ag}.  This model, however, could
not reproduce the central value of the running spectral index obtained
by WMAP data but could realize the feature only at the one-sigma level
because small-scale fluctuations tend to be too large and we have to
hide the corresponding scales in an unobservable region.  More serious
is the problem of the initial condition common to other hybrid inflation
models that only a very limited initial configuration can lead to
inflation \cite{hybridinit}.  Both these problems have been solved in
the chaotic hybrid new inflation model in supergravity proposed by us
\cite{YY}, which can also realize mildly large fluctuations in the
appropriate scales to realize early star formation to help early
re-ionization.

In the present paper we present another possible mechanism to realize
running spectral index from $\ns >1$ to $\ns < 1$ for decreasing length
scale: namely, smooth hybrid inflation in supergravity.  This scenario
was originally proposed by Lazarides and Panagiotakopoulos
\cite{smooth,shift,SS}, in which nonrenormalizable terms are introduced
and gauge symmetry remains broken even during hybrid inflation. Thus,
topological defects are not produced at the end of inflation. In this
paper, we discuss smooth hybrid inflation in supergravity and
investigate whether the running spectral index is obtained with the
desired property.  As will be shown shortly, the spectral index runs
from $\ns>1$ on a large scale to $\ns<1$ on a smaller scale without
resorting to the one-loop effects unlike our previous models
\cite{KYY,YY}.  Another merit of this scenario is that it can be
realized with natural initial conditions in minimal supergravity
\cite{smoothini}.
%The simple sum of
%one-loop effects and supergravity effects is not necessarily
%justified. 
We find, however, that we cannot yield large enough $e$-folds of
inflation with large enough running of the spectral index whichever
power of the nonrenormalizable term we may choose.  So another inflation
is required after smooth hybrid inflation as with the case with our
previous models \cite{YY,KYY}.  Adopting new inflation as the second
inflation, we make a specific model of smooth hybrid new inflation in
supergravity.  Generally speaking, density fluctuations produced at the
onset of new inflation become large.  Actually, as shown in \cite{KYY},
if we consider usual hybrid inflation before new inflation, the density
fluctuations produced during new inflation are too large, which may
cause an overproduction problem of dark halos.  However, in the case of
smooth hybrid inflation, they can be adequately large, which may be
helpful for early star formation.  Thus this scenario can also solve the
two problems of the hybrid new inflation model of \cite{KYY} just as the
chaotic hybrid new inflation model \cite{YY} does.  Which of these two
remaining models is more appropriate may be judged by future
observations from the presence or the absence of cosmic strings, because
the latter model predicts cosmic strings with their energy scale close
to the current observational upper bound imposed by cosmic microwave
background radiation.  On the other hand, it has been claimed that long
cosmic strings lose their energy directly into particles instead of
string loops \cite{Vincent:1996rb}.  Although we understand this issue
is still in dispute \cite{moore}, if it turned out to be true, it would
rule out our previous model because too many high-energy cosmic rays
would be produced \cite{Wichoski:1998kh}.

The rest of the paper is organized as follows. In Sec. II we consider
smooth hybrid inflation in supergravity and investigate the spectral
nature of produced density fluctuations. Then in Sec. III, after
reviewing new inflation, we introduce smooth hybrid new inflation, and
investigate their dynamics and density fluctuations. Section IV is
devoted to a discussion and future outlook. In this paper, we set
$\mg=2.4\times 10^{18}$ GeV to be unity otherwise stated.
 
%%%%%%%%%%%%%%%%%%%%%%%%%%%%%%%%%%%%%%%%%%%%%%%%%%%%%%%%%%%%%%%%%%%%%
\section{Smooth hybrid inflation in supergravity}
%%%%%%%%%%%%%%%%%%%%%%%%%%%%%%%%%%%%%%%%%%%%%%%%%%%%%%%%%%%%%%%%%%%%%

\label{sec:hybrid}

First we introduce smooth hybrid inflation in supergravity
\cite{smooth,SS}. The superpotential is given by
\beq
  W_H= S \lmk -\mu^2 + \frac{(\Psibar\Psi)^m}{M^{2(m-1)}} \rmk, 
  \label{Hsuper}
\eeq
where $\Psi$ and $\Psibar$ are a conjugate pair of superfields
transforming as nontrivial representations of some gauge group, under
which a superfield $S$ is singlet. $\mu$ is the energy scale of hybrid
inflation which may be related to the grand unified theory, and $M$ is a
cutoff scale which controls the nonrenormalizable term.  $m$ is an
integer with $m \geq 2$.  This superpotential possesses two
symmetries. One is the $R$ symmetry under which they are transformed as
$S\longrightarrow e^{2i\alpha}S$, $\Psi \longrightarrow
e^{2i\alpha}\Psi$, $\Psibar \longrightarrow e^{-2i\alpha}\Psibar$, and
$W_H \longrightarrow e^{2i\alpha}W_H$. The other is a discrete symmetry
$Z_m$ under which $\Psibar\Psi$ has unit charge. The $R$-invariant
K\"ahler potential is given by
\beq
 K_H=|S|^2+|\Psi|^2+|\Psibar|^2+\cdots, 
\label{Hkaehler}
\eeq
where we have retained only the minimal terms.  But the essential result
remains intact even if we take higher-order terms into account.

The potential of scalar fields is given by
\bea
  V_H(S,\Psi,\Psibar) &=& e^{|S|^2+|\Psi|^2+|\Psibar|^2}
                      \lnk (1-|S|^2+|S|^4)
              \left| -\mu^2 + \frac{(\Psibar\Psi)^m}{M^{2(m-1)}} \right|^2 
                      \right. \non \\
                      && + |S|^2 \lkk \left|
                          m\Psibar\frac{(\Psibar\Psi)^{m-1}}{M^{2(m-1)}}  
                          +\Psi^\ast 
              \lmk -\mu^2 + \frac{(\Psibar\Psi)^m}{M^{2(m-1)}} \rmk
                                             \right|^2 \right. \non \\
                      && \qquad\,\, \left. \left.
                                           + \left|
                          m\Psi\frac{(\Psibar\Psi)^{m-1}}{M^{2(m-1)}}  
                          +\Psibar^\ast 
              \lmk -\mu^2 + \frac{(\Psibar\Psi)^m}{M^{2(m-1)}} \rmk
                                             \right|^2  
                                         \rkk \rnk + V_D,
\eea
where the scalar components of the superfields are denoted by the same
characters as the corresponding superfields. Here $V_{D}$ represents the
$D$-term contribution and is given by
\beq
  V_{D} = \frac{e^2}{2} (|\Psi|^2 - |\overline{\Psi}|^2)^2,
\eeq 
where we assumed for simplicity that the gauge group $G$ is U$(1)$ and
$e$ is the gauge coupling constant. Then, the $D$-term contribution
vanishes for the direction $|\Psi| = |\overline{\Psi}|$. On the other
hand, the steepest descent direction in the $F$-term contribution is
$(\Psibar\Psi)^m = |\Psibar\Psi|^m$, which is compatible with the
$D$-flat condition. Performing adequate gauge, discrete, and $R$
transformations, the complex scalar fields are changed into real scalar
fields, $\sigma \equiv \sqrt{2}\,\rm{Re}\,S$, $\psi \equiv
2\,\rm{Re}\,\Psi = 2\, \rm{Re}\,\Psibar$. Then, neglecting higher-order
terms, the scalar potential is given by
\bea
  V_H(\sigma,\psi) &\cong& \lmk 
                         -\mu^2 + \frac{\psi^{2m}}{4^m M^{2(m-1)}}
                        \rmk^2
                        \lmk 1 + \frac{\psi^2}{2} + \frac18\,\sigma^4
                        \rmk \non \\
                   &&   +    \frac{m^2}{4^{2m-1}}\,\sigma^2\psi^2 
                         \lmk \frac{\psi}{M} \rmk^{4(m-1)} 
                        -\frac{m\mu^2\psi^{2m}\sigma^2}
                       {2^{2m-1}M^{2(m-1)}}
\eea
in the regime $|\sigma| \ll 1$ and $|\psi| \ll 1$.
The minimum of $\psi$ is estimated as
\beq
  \psi_{\rm min} \cong \lhk
     \bear{ll}
      \displaystyle{\lmk \frac{4^m}{2m(2m-1)} \rmk^{\frac{1}{2(m-1)}}
                          M \lmk \frac{\mu}{\sigma}
                             \rmk^{\frac{1}{m-1}} } &
            \qquad \mbox{for $\sigma \gg (\mu M^{m-1})^{\frac1m}$}, 
            \\ [0.5cm]
      \displaystyle{2 (\mu M^{m-1})^{\frac1m} } &
            \qquad \mbox{for $\sigma \ll (\mu M^{m-1})^{\frac1m}$}.
     \eear \right.
\eeq
Then, the effective potential of $\sigma$ is given by
\beq
  V_H(\sigma) \cong \mu^4 \lkk 1
                      - \frac{4^\frac{2m-1}{m-1}}
                             {[2m(2m-1)]^\frac{m}{m-1}}
                           \lmk \frac{m-1}{4m-2} \rmk M^2 
                          \lmk \frac{\mu}{\sigma^m}
                          \rmk^{\frac{2}{m-1}}
                      + \frac{1}{8} \sigma^4 + \cdots \rkk
\eeq     
for $\sigma \gg (\mu M^{m-1})^{\frac1m}$. Here we have set $\psi =
\psi_{\rm min} \ne 0$, which breaks the gauge symmetry so that no
topological defect is formed at the termination of
inflation.\footnote{Since the effective mass squared in the direction of
$\psi$ is much larger than the Hubble squared, $\psi$ quickly traces
$\psi_{\rm min}$. However, such an approximation still may cause small
errors to the estimates of an $e$-fold number, density fluctuations, and
so on.} As long as $\sigma$ is smaller than unity, the effective
potential is dominated by the false vacuum energy $\mu^4$.

The derivative of the effective potential is given by
\beq
  \frac{dV_H}{d\sigma} \cong \mu^4 \lkk
                      \frac{4^\frac{2m-1}{m-1}}
                             {[2m(2m-1)]^\frac{m}{m-1}}
                             \lmk \frac{m}{2m-1} \rmk M^2 
                          \lmk \frac{\mu^2}{\sigma^{3m-1}}
                          \rmk^{\frac{1}{m-1}}
                      + \frac{1}{2} \sigma^3 + \cdots \rkk.
\eeq     
The dynamics is determined by the first term for $\sigma < \sigma_d$ and
the last term for $\sigma > \sigma_d$ with
\beq
  \sigma_d \equiv \lmk \frac{2m}{2m-1} \rmk^{\frac{m-1}{6m-4}}
                  \lmk \frac{4^{2m-1}}{[2m(2m-1)]^m}
                    \rmk^{\frac{1}{6m-4}}
                  \lmk \mu M^{m-1} \rmk^{\frac{1}{3m-2}}.
  \label{eq:sigmad}
\eeq   
The slow-roll condition is broken and inflation ends at $\sigma \simeq
\sigma_c$ when $3H^2 = |d^2V_H/d\sigma^2|$. Here $\sigma_c$ is given by
\beq
  \sigma_c \equiv \lmk \frac{m}{2m-1}\,\frac{3m-1}{m-1} \rmk^{\frac{m-1}{4m-2}}
                  \lmk \frac{2}{[2m(2m-1)]^{\frac{m}{4m-2}}}
                    \rmk
                  \lmk \mu M^{m-1} \rmk^{\frac{1}{2m-1}}.
\eeq   
Since $\mu \ll1 $ and $ M \lesssim 1$, we find $\sigma_d \gg
\sigma_c$.  Then, the number of $e$-folds of smooth hybrid inflation,
$N_H$, is estimated as
\bea
  N_H &=& \int^{\sigma_i}_{\sigma_c} d\sigma \frac{V_H}{V_H'} 
    \sim \int^{\sigma_i}_{\sigma_d} d\sigma \frac{2}{\sigma^3}
        + \int^{\sigma_d}_{\sigma_c} d\sigma 
               \frac{[2m(2m-1)]^\frac{m}{m-1}}{4^\frac{2m-1}{m-1}}
                  \lmk \frac{2m-1}{m} \rmk
                    \frac{\sigma^\frac{3m-1}{m-1}}
                         {\mu^{\frac{2}{m-1}}M^2} \non \\
    &\sim& \lmk \frac{3m-2}{2m-1} \rmk 
         \lmk \frac{2m-1}{2m} \rmk^\frac{m-1}{3m-2}
          \frac{[2m(2m-1)]^\frac{m}{3m-2}}{4^\frac{2m-1}{3m-2}}
           \lmk \mu M^{m-1} \rmk^{-\frac{2}{3m-2}},
\eea
where $\sigma_i$ is an initial value of inflaton $\sigma$ and the prime
represents the derivative with respect to $\sigma$. In the large $m$
limit, $N_H$ is proportional to $(m/M)^{2/3}$ asymptotically.

We define the slow-roll parameters to obtain the density fluctuations
produced during smooth hybrid inflation:
\bea
  \epsilon_H &\equiv& \frac12 \lmk \frac{V_H'}{V_H} \rmk^2
           = \frac18 \lkk \sigma_d^3 
                 \lmk \frac{\sigma_d}{\sigma} \rmk^\frac{3m-1}{m-1}
                        + \sigma^3
                     \rkk^2 = \CO ( \sigma^6 ), \non \\
  \eta_H &\equiv& \frac12 \lmk \frac{V_H''}{V_H} \rmk
           = - \frac12 \lmk \frac{3m-1}{m-1} \rmk \sigma_d^2 
                 \lmk \frac{\sigma_d}{\sigma} \rmk^\frac{4m-2}{m-1}
                        + \frac32 \sigma^2
                            = \CO ( \sigma^2), \non \\ 
  \xi_H &\equiv& \frac12 \lmk \frac{V_H'''V_H'}{V_H^2} \rmk
           = \frac{1}{2} \lkk 
               \frac{(3m-1)(2m-1)}{(m-1)^2}\,\sigma_d 
                 \lmk \frac{\sigma_d}{\sigma} \rmk^\frac{5m-3}{m-1}
                 + 3\sigma \rkk
                          \lkk \sigma_d^3 
                 \lmk \frac{\sigma_d}{\sigma} \rmk^\frac{3m-1}{m-1}
                 + \sigma^3 \rkk \non \\
      &=& \CO (\sigma^4).
\eea 
By using these parameters, the spectral index of density fluctuations
$n_s$ and its derivative $dn_s/d\ln k$ are evaluated as \cite{LL}
\bea
  n_s - 1 &=& - 6\epsilon_H + 2\eta_H \cong 2 \eta_H, \non \\ 
  \frac{dn_s}{d\ln k} &=& 16\epsilon_H\eta_H - 24\epsilon_H^2 -2\xi_H
           \cong -2\xi_H.
  \label{eq:index}
\eea
Note that $n_s = 1$ at $\sigma = [(3m-1)/(3m-3)]^{(m-1)/(6m-4)}
\sigma_d < \sigma_d$. Conversely, given $n_s$ and $dn_s/d\ln k$, we can
obtain $\sigma$ and $\sigma_d$ from the above equation:
\bea
  \sigma &=& \frac{1}{2^{\frac34} \sqrt{3(3m-2)}} \lkk
            \lmk 25m^2-34m+13 \rmk \lmk n_s-1 \rmk^2 
           - 12 \lmk 3m-1 \rmk \lmk m-1 \rmk \frac{dn_s}{d\ln k}
           \right. \non \\
         &&\left. + \lmk 7m-5 \rmk \lmk n_s-1 \rmk
             \sqrt{\lmk m+1 \rmk^2 \lmk n_s-1 \rmk^2
                   - 24 \lmk 3m-1 \rmk \lmk m-1 \rmk \frac{dn_s}{d\ln k}
                  }
                                               \,\rkk^{\frac14}, \non \\
  \sigma_d &=& \lkk \frac{m-1}{3m-1} (3\sigma^2 - n_s + 1)
               \,\rkk^\frac{m-1}{6m-4}
               \sigma^\frac{2m-1}{3m-2}.
\eea
On the other hand, the amplitude of of curvature perturbation in the
comoving gauge $\CR$ is given by \cite{pert}
\beq
  \CR \cong \frac{1}{2\pi} \sqrt{\frac{V_H}{6\epsilon_H}}
      \cong \frac{\mu^2}{\sqrt{3}\pi}
              \lkk \sigma_d^3 
                 \lmk \frac{\sigma_d}{\sigma} \rmk^\frac{3m-1}{m-1}
                        + \sigma^3
                     \rkk^{-1}.
\eeq
Inserting the central value $\CR = 4.7 \times 10^{-5}$ 
%corresponding to $A = 0.75$ 
on the comoving scale $k_0=0.002\,\,Mpc^{-1}$, which is obtained by
WMAPext+2dFGRS+Ly$\alpha$ \cite{Peiris}, the energy scale $\mu$ is given
by
\beq
  \mu = 1.6 \times 10^{-2} \sqrt{\sigma_d^3 
                 \lmk \frac{\sigma_d}{\sigma} \rmk^\frac{3m-1}{m-1}
                        + \sigma^3}.
\eeq
Using Eq.\ (\ref{eq:sigmad}), we can also evaluate the other energy
scale $M$ from $\sigma_d$ and $\mu$.

\begin{figure}
\includegraphics[width=16cm]{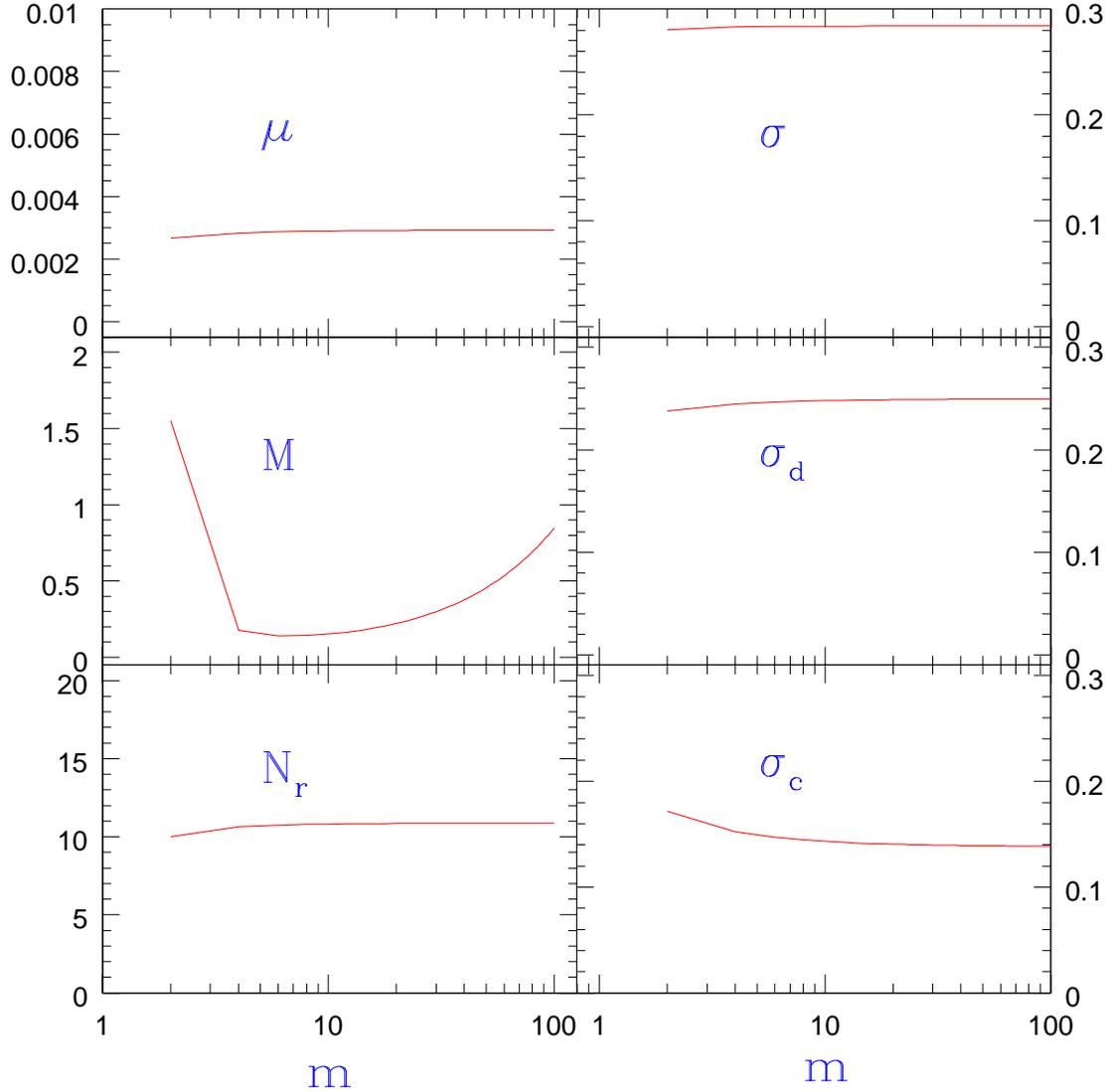} 
\caption{\label{fig:para} 
  The values of all parameters are shown for each $m$.}
\end{figure}

Inserting $n_s = 1.13$ and $dn_s/d\ln k = -0.055$ on the comoving scale
$k_0=0.002\,\Mpc^{-1}$ obtained by WMAPext+2dFGRS+Ly$\alpha$
\cite{Peiris}, we can obtain the values of all parameters for each
$m$. For example, in the case $m=2$, $\mu = 2.7 \times 10^{-3} = 6.5
\times 10^{15}$ GeV, $M = 1.6, \sigma = 0.28$, and $\sigma_d = 0.24$.
The number of $e$-folds, $N_r$, of smooth hybrid inflation after the
scale with the desired spectral shape has crossed the Hubble radius is
found to be 10.  Also, in the case of $m=8$, $\mu = 2.9 \times 10^{-3} =
7.0 \times 10^{15}$ GeV, $M = 0.14, \sigma = 0.28, \sigma_d = 0.25$, and
$N_r = 11$. The values of parameters for each $m$ is depicted in Fig.\
\ref{fig:para}. Note that even if we take large $m$, $N_r$ does not
increase in proportion to $m^{2/3}$, because an increase in $m$ is
accompanied by that in $M$ as long as we use the observed values of
$\ns,~dn_s/d\ln k$, and $\CR$.  As a result $N_r$ does not change
significantly no matter how large $m$ may be, as seen in Fig.\
\ref{fig:para}.  Therefore, another inflation is necessary in order that
the scale with the desired spectral shape be pushed to the comoving
scale $k_0=0.002\,\Mpc^{-1}$.

%%%%%%%%%%%%%%%%%%%%%%%%%%%%%%%%%%%%%%%%%%%%%%%%%%%%%%%%%%%%%%%%%%%%%
\section{Smooth hybrid new inflation in supergravity}
%%%%%%%%%%%%%%%%%%%%%%%%%%%%%%%%%%%%%%%%%%%%%%%%%%%%%%%%%%%%%%%%%%%%%

\label{sec:hybridnew}

As the second inflation to follow smooth hybrid inflation, we adopt new
inflation proposed by Izawa and Yanagida \cite{IY}. While hybrid
inflation including smooth hybrid inflation predicts a high reheating
temperature because the inflaton has gauge couplings and its energy
scale is relatively high usually, new inflation predicts a sufficiently
low reheating temperature to avoid overproduction of gravitinos. Thus,
the occurrence of new inflation following smooth hybrid inflation is
favorable also in this respect. Furthermore, in our model, the initial
value of new inflation is dynamically set during smooth hybrid
inflation, which evades the severe initial value problem of new
inflation. In this section, after reviewing new inflation briefly, we
discuss smooth hybrid new inflation in supergravity by considering full
superpotential and K\"ahler potential.

The superpotential of new inflation is given by
\beq
 W_N[\Phi] = v^2\Phi - \frac{g}{n+1}\Phi^{n+1}, \label{Nsuper}
\eeq
where we introduce a chiral superfield $\Phi$ with an $R$ charge
$2/(n+1)$, but assume that the U$(1)_R$ symmetry is dynamically broken
to a discrete $Z_{2n~R}$ symmetry at a scale $v \ll \mu$. Here $g$ is a
coupling constant of order of unity and we also assume that both $g$ and
$v$ are real and positive for simplicity.  We assume $n$ is larger than
2 because $g$ must be unnaturally small to realize inflation if $n=2$.
The $R$-invariant K\"ahler potential is given by
\beq
 K_N=|\Phi|^2+\frac{\kn}{4}|\Phi|^4+\cdots, \label{Nkaehler}
\eeq
where $\kn$ is a constant smaller than unity.  

From Eqs. (\ref{Nsuper}) and (\ref{Nkaehler}), the scalar potential of
new inflaton reads
\bea
V_N[\Phi]&=&\frac{\exp\lmk |\Phi|^2+\frac{\kn}{4}|\Phi|^4\rmk}
{1+\kn|\Phi|^2} \nonumber \\
& &\times\lkk~ \left| \lmk 1+|\Phi|^2+\frac{\kn}{2}|\Phi|^4\rmk v^2
-\lmk 1+\frac{|\Phi|^2}{n+1}+\frac{\kn|\Phi|^4}{2(n+1)}\rmk
g\Phi^n \right|^2 \right. \nonumber \\ 
& &~~~\left.
-3\lmk 1+\kn|\Phi|^2 \rmk|\Phi|^2\left| v^2-\frac{g}{n+1}\Phi^n
\right|^2\rkk .
\eea
It has a minimum at
\beq
 |\Phi|_{\min}\cong\lmk\frac{v^2}{g}\rmk^{\frac{1}{n}}
 ~~~{\rm and~~Im}\Phi^n_{\min}=0,
\eeq
with a negative energy density
\beq
 V_N[\Phimin]\cong -3e^{K_N}|W_N[\Phimin]|^2\cong
 -3\lmk\frac{n}{n+1}\rmk^2v^4|\Phimin|^2.
\eeq
As was done in \cite{IY}, we assume that this negative energy density is
canceled by a positive contribution coming from supersymmetry breaking,
$\Lambda_{\rm SUSY}^4$, which relates energy scale $v$ of this model
with the gravitino mass $m_{3/2}$ as
\beq
 m_{3/2}\cong \frac{n}{n+1}\lmk\frac{v^2}{g}\rmk^{\frac{1}{n}}v^2.
\eeq

Then, the potential of new inflaton is approximated as
\beq
 V_N[\phi]\cong v^4-\frac{\kn}{2}v^4\phi^2-\frac{2g}{2^{n/2}}v^2\phi^n 
 +\frac{g^2}{2^n}\phi^{2n}, \label{Neffpote}
\eeq
where we identified the real part of $\Phi$ with the inflaton
$\phi \equiv \sqrt{2}\,\rm{Re}\,\Phi$.
The derivative of the effective potential is given by
\beq
  \frac{dV_N}{d\phi} \cong -\kn v^4\phi 
      -{2^{\frac{2-n}{2}}}{ngv^2} \phi^{n-1}
      + 2^{1-n} n g^2 \phi^{2n-1}.
  \label{Neqm}
\eeq     
Since the last term is negligible during inflation, the dynamics is
determined by the first term for $\phi < \phi_d$ and the second term for
$\phi > \phi_d$ with
\beq
  \phi_d \equiv \sqrt{2}\lmk\frac{\kn v^2}{gn}\rmk^{\frac{1}{n-2}}.
\eeq
Then, the slow-roll parameters are given by
\beq
 \epsilon_N \cong \frac{1}{2}\lmk\kn\phi+{2^{\frac{2-n}{2}}}{ng}
                    \frac{\phi^{n-1}}{v^2}\rmk^2,~~~
 \eta_N = -\kn-{2^{\frac{2-n}{2}}}{n(n-1)g}\frac{\phi^{n-2}}{v^2}.
\eeq
Thus inflation is realized for $\kn \ll 1$ and ends at $\phi = \phi_c$
defined as
\beq
  \phi_c \equiv
       \sqrt{2}\lmk\frac{(1-\kn)v^2}{gn(n-1)}\rmk^{\frac{1}{n-2}},
\eeq
when $|\eta|$ becomes as large as unity.

The total number of $e$-folds of new inflation is estimated as
\beq
 N_N=-\int_{\phi_i}^{\phi_c} d\phi \frac{V_N}{dV_N/d\phi}
 \cong\int_{\phi_i}^{\phi_d} d\phi \frac{d\phi}{\kn\phi} +
\int_{\phi_d}^{\phi_c}\frac{2^{\frac{n-2}{2}}v^2}{gn\phi^{n-1}}
=\frac{1}{\kn}\ln\frac{\phi_d}{\phi_i}+\frac{1-n\kn}{(n-2)\kn(1-\kn)}
 \label{nn} 
\eeq
for $0<\kn<n^{-1}$. If $\kn$ vanishes, we instead find
\beq
 N_N=\int_{\phi_i}^{\phi_c} d\phi
\frac{2^{\frac{n-2}{2}}v^2}{gn\phi^{n-1}}
=\frac{2^{\frac{n-2}{2}}v^2}{gn(n-2)}
\phi_i^{2-n}-\frac{n-1}{n-2}. \label{nnzero}
\eeq
Here $\phi_i$ is the initial value of $\phi$, which is set dynamically
during smooth hybrid inflation, as shown below.

We investigate full dynamics of smooth hybrid new inflation by assuming
that there are no direct interactions between fields relevant to hybrid
inflation discussed in the previous section and $\Phi$. That is, the
full superpotential and K\"ahler potential is given by $W=W_H+W_N$ and
$K=K_H+K_N$.

In the hybrid inflation stage $\sigma > \sigma_c$, the cosmic energy
density is dominated by the false vacuum energy $\mu^4$, which gives the
interaction terms between $S$ and $\Phi$:
\beq
 V \supset \mu^4|\Phi|^2+\muni v^2 (\Phi^\ast S+\Phi S^\ast)+\cdots
=\frac{1}{2}\mu^4(\phi^2+\chi^2)+\muni
v^2\sigma\phi+\cdots.
\eeq
Hence at the end of hybrid inflation, $\sigma=\sigma_c$, $\phi$ and
$\chi \equiv \sqrt{2}\,\rm{Im}\,\Phi$ have a minimum at
\beq
 \phi_{\min}\cong-\frac{v^2}{\muni}\sigma_c
 = - 5^{\frac{1}{6}}\lmk\frac{2}{3}\rmk^{\frac{1}{2}} \frac{v^2}{\mu^2}
     (\mu M)^\frac13, 
~~~~~~ \chi_{\min}\cong 0,
\label{eq:minimum}
\eeq
respectively.
Here and hereafter we set $m=2$ for definiteness.

Since the effective mass is larger than the Hubble parameter during
hybrid inflation, the above configuration is realized with the
dispersion
\beq
 \langle(\phi-\phi_{\min})^2\rangle=\langle\chi^2\rangle
=\frac{3}{8\pi^2}\frac{H_H^4}{\mu^4}=\frac{\mu^4}{24\pi^2},
\eeq
due to quantum fluctuations \cite{BD}, where $H_H\equiv \mu^2/\sqrt{3}$
is the Hubble parameter during smooth hybrid inflation. The ratio of
quantum fluctuation to the expectation value should satisfy
\beq
 \frac{\sqrt{\langle(\phi-\phi_{\min})^2\rangle}}{|\phi_{\min}|}
=\frac{1}{4 \times 5^{\frac{1}{6}}\pi} 
  \lmk\frac{\mu^5}{M}\rmk^{\frac13} \lmk \frac{\mu}{v} \rmk^2
\ll 1, 
\label{ratio}
\eeq
so that the initial value of the inflaton for new inflation is located
off the origin with an appropriate magnitude.

After $\sigma$ reaches $\sigma_c$, smooth hybrid inflation ends and the
fields $S$ and $\Psi (\Psibar)$ start oscillating and decay eventually.
If this oscillation phase lasts for a prolonged period due to
gravitationally suppressed interactions of these fields, $\phi$ will
also oscillate and its amplitude decreases with an extra factor $v/\mu$
\cite{Kawasaki:1998vx}.  In this case, new inflation could start with an
even smaller value of $\phi$ depending on its phase of oscillation at
the onset of inflation (see \cite{Kanazawa:1999ag} for an analytic
estimate of the initial phase). So we set the initial value of $\phi$ as
\beq
\phi_i = \lmk \frac{v}{\mu} \rmk \phi_{\min},
\eeq
and new inflation occurs until $\phi=\phi_c$ with the potential
(\ref{Neffpote}). 

Contrary to the smooth hybrid inflation regime we do not have much
precise observational constraints on the new inflation regime, so we
cannot fully specify values of the model parameters for new
inflation. Hence let us content ourselves with a few specific
examples. First we consider the cases with $\kn=0$. Then from
Eq. (\ref{nnzero}) the number of $e$-folds of new inflation reads
\beq
 N_N=\frac{2^{\frac{n-2}{2}} v^2}{gn(n-2)} \phi_i^{2-n}
-\frac{n-1}{n-2}
\cong \frac{2^{\frac{n-2}{2}} v^2}{gn(n-2)} \phi_i^{2-n}. \label{nnn}
\eeq
This should be around $40 + (2/3)\ln(\mu/v)$ to push the comoving scale
with appropriate spectral shape to the appropriate physical length
scale.\footnote{
  Comoving scales that left the Hubble radius in the late stage of
 hybrid inflation reenter the horizon before the beginning of the new
 inflation. Hence extra $e$-folds $(2/3)\ln(\mu/v)$ should be added in
 making a correspondence between comoving horizon scales during hybrid
 inflation and proper scales \cite{Kawasaki:1998vx}.  }
On the other hand, the amplitude of curvature perturbation at the onset
of new inflation, $\phi=\phi_{i}$, is given by
\beq
  \calr = \frac{n-2}{2\sqrt{3}\pi} N_N \frac{\mu^3}{v\sigma_c}
        \cong 1.5 \times 10^{-4} (n-2)\frac{\mu}{v}\frac{N_N}{40},
\eeq
where use has been made of the values $\mu=2.7\times 10^{-3}$ and
$\sigma_c=0.17$ in the last equality. From Eq. (\ref{nn})
we find
\bea
 v&=&9.4\times 10^{-5}\lmk\frac{gN_N}{40}\rmk^{-\frac{1}{4}},~~~{\rm
for}~n=4, \nonumber \\
 v&=&9.6\times 10^{-4}\lmk\frac{gN_N}{40}\rmk^{-\frac{1}{10}},~~~{\rm
for}~n=6, \nonumber \\
 v&=&1.8\times 10^{-3}\lmk\frac{gN_N}{40}\rmk^{-\frac{1}{16}},~~~{\rm
for}~n=8, \nonumber \\
 v&=&2.3\times 10^{-3}\lmk\frac{gN_N}{40}\rmk^{-\frac{1}{22}},~~~{\rm
for}~n=10, \nonumber \\
 v&=&2.7\times 10^{-3}\lmk\frac{gN_N}{40}\rmk^{-\frac{1}{28}},~~~{\rm
for}~n=12.
\eea
For $g<1$ and $n\geq 12$, $v$ is larger than $\mu$, which contradicts
with our assumption that new inflation takes place after hybrid
inflation at lower energy scale. Setting $g=1$, we find for $n=4$ that
$\CR \cong 8.7 \times 10^{-3}$ at the onset of new inflation which
corresponds to the comoving scale $\ell_{*}\sim 1.3$ Mpc today. $v$
can be close to $\mu$ and $\CR$ can be smaller only for $g \lesssim
10^{-6}$, which is quite unnatural.  For $n=8$, we find that $\CR
\cong 1.4 \times 10^{-3}$ at the onset of new inflation which
corresponds to the comoving scale $\ell_{*}\sim 190$ kpc today. These
fluctuations cause the early formation of dark halo objects with
comoving scale $\ell_{*}$.  In the above cases, since $\ell_{*}$ is
larger than about 1~kpc, the dark halos may cause a cosmological
problem because they significantly harm subsequent galaxy formation or
produce too many gravitational lens events.

On the other hand, for $\kn\neq 0$, we find
\beq
 \calr =\frac{\mu^2}{2\sqrt{3}\pi\kn \sigma_c} \frac{\mu}{v}
      \cong 3.8\times 10^{-6} \kn^{-1}\frac{\mu}{v},
\eeq
at the onset of new inflation.  It is independent on $n$ and $g$.
% and is a little smaller than $10^{-5}$. 
From Eq. (\ref{nn}), $v$ is related to
the number of $e$-folds $N_N$:
\beq
  v = \lkk 
         \frac{\sigma_c}{\sqrt{2}\mu^3} 
         \lmk \frac{\kn}{ng} \rmk^{-\frac{1}{n-2}}
         \exp{\lhk \kn \lmk N_N - \frac{1-n\kn}{(n-2)\kn(1-\kn)} \rmk 
              \rhk} 
      \rkk^{\frac{n-2}{8-3n}}. 
\eeq
Thus, if we take $\kn=0.1$, for example, $N_N=40$ implies
\bea
 v&=&2.5\times 10^{-5} g^{-\frac{1}{4}},~~~{\rm
for}~n=4, \nonumber \\
 v&=&2.7\times 10^{-4} g^{-\frac{1}{10}},~~~{\rm
for}~n=6, \nonumber \\
 v&=&4.8\times 10^{-4} g^{-\frac{1}{16}},~~~{\rm
for}~n=8, \nonumber \\
 v&=&6.4\times 10^{-4} g^{-\frac{1}{22}},~~~{\rm
for}~n=10.
\eea
For $n=4$, $\CR \cong 4.1 \times 10^{-3} g^{1/4}$ at the comoving
scale $\ell_{*}\cong 3.1 g^{1/6}$~Mpc, which again causes the dark
halo problem unless $g$ is extremely small. On the other hand, for $n
\ge 6$ and $g=1$, $\CR \lesssim 3.8 \times 10^{-4}$ at the comoving
scale $\ell_{*} \lesssim 650$~kpc, which may be helpful for early star
formation which is required for early re-ionization \cite{Bennett} and
from the age estimate of high-redshift quasars using the cosmological
chemical clock \cite{jy}.

%%%%%%%%%%%%%%%%%%%%%%%%%%%%%%%%%%%%%%%%%%%%%%%%%%%%%%%%%%%%%%%%%%%%%
\section{Discussion and Conclusion}
%%%%%%%%%%%%%%%%%%%%%%%%%%%%%%%%%%%%%%%%%%%%%%%%%%%%%%%%%%%%%%%%%%%%%

In this paper we proposed a new model of inflation in supergravity, in
which the two new features discovered by the recent precision
measurements of cosmic microwave background anisotropy can be explained
simultaneously and naturally: namely, the running of spectral index of
density fluctuations on large scale as preferred by the first-year WMAP
data and a large enough amplitude of fluctuation on small scale relevant
to first star formation to realize early re-ionization as discovered by
WMAP.  The desired running feature--that is, the spectral index with
$\ns >1$ on a large scale and $\ns <1$ on a smaller scale--is naturally
generated by the balance between the nonrenormalizable term in the
superpotential and supergravity effects without resorting to the
one-loop effect contrary to our previous models \cite{YY,KYY}.

Compared with the chaotic hybrid new inflation model in supergravity
\cite{YY}, the present model has somewhat simpler symmetry structures,
although we have been unable to explain the hierarchy of the energy
scales of two inflations here unlike in the previous model \cite{YY}.
Because our previous model induces string formation with a fairly
large energy scale, if forthcoming analysis could rule out such
topological defects, the present model would be the only surviving
model among the two.

\acknowledgments{ J.Y.\ is grateful to Robert H. Brandenberger for his
hospitality at Brown University, where this work started. We thank
V. N. Senoguz for useful comments. This work was partially supported by
JSPS Grant-in-Aid for Scientific Research No.\ 13640285 (J.Y.) and the
JSPS for research abroad (M.Y.).  M.Y.\ is partially supported by the
Department of Energy under Grant No.\ DEFG0291ER40688.}

\end{document}